**RESEARCH ARTICLE**

# Sensitivity analysis of G-estimators to invalid instrumental variables

## Valentin Vancak | Arvid Sjölander

Dept. of Medical Epidemiology and
Biostatistics, Karolinska Institutet,
Stockholm, Sweden

**Correspondence**
Valentin Vancak, Dept. of Medical
Epidemiology and Biostatistics,
Karolinska Institutet, Stockholm,
Sweden.
Email: valentin.vancak@ki.se

**Funding information**
Vetenskapsrådet, Grant/Award Number:
2020-01188

**Abstract**

Instrumental variables regression is a tool that is commonly used in the analysis of observational data. The instrumental variables are used to make causal inference about the effect of a certain exposure in the presence of unmeasured confounders. A valid instrumental variable is a variable that is associated with the exposure, affects the outcome only through the exposure (exclusion), and is not confounded with the outcome (exogeneity). Unlike the first assumption, the other two are generally untestable and rely on subject-matter knowledge. Therefore, a sensitivity analysis is desirable to assess the impact of assumptions' violation on the estimated parameters. In this paper, we propose and demonstrate a new method of sensitivity analysis for G-estimators in causal linear and non-linear models. We introduce two novel aspects of sensitivity analysis in instrumental variables studies. The first is a single sensitivity parameter that captures violations of exclusion and exogeneity assumptions. The second is an application of the method to non-linear models. The introduced framework is theoretically justified and is illustrated via a simulation study. Finally, we illustrate the method by application to real-world data and provide guidelines on conducting sensitivity analysis.

## 1 | INTRODUCTION

In many studies, the goal is to estimate the causal effect of a certain exposure $X$ on an outcome $Y$. The estimation is straightforward when a randomized controlled trial (RCT) with perfect compliance is feasible. The randomization ensures that, in large samples and with full compliance, the only systematic difference between the two groups is the exposure itself. Therefore, any observed exposure-outcome association, for example, the risk difference (RD) and the odds ratio (OR), can be interpreted as a causal effect.\*Formally, an RCT allows us to infer the counterfactual probability of the outcome had everyone, contrary to fact, been exposed (unexposed). This is in contrast to RCTs with imperfect compliance, where the differences between exposed and unexposed cannot be attributed only to the exposure effect. Therefore, the causal effect of interest cannot be easily computed using naive statistical association measures. However, in such a situation, one can still identify the causal effect of interest by using the original assignment $Z$. The original assignment is associated monotonically with the exposure (relevance), affects the outcome only through the exposure (exclusion), and is not confounded with the outcome (exogeneity). Notably, these three properties define a valid instrumental variable (IV). Such variables allow for the identification of causal effects, particularly the local average treatment effect (LATE), which is the average treatment effect (ATE) for individuals whose treatment status can be influenced by changing the IV value.[2] These causal effects are the target effects in many observational studies and the main focus of this article.

Contrary to the RCTs, finding a valid IV in observational studies is complicated since the exclusion and the exogeneity properties are testable only under strict constraints.[3] For example, Palmer et al[4] developed a statistical test based on inequalities induced on the joint distribution of the IV, the exposure, and the outcome. However, this test is only applicable to categorical data, which is an important limitation. Another approach utilizes the overidentification. Overidentification refers to a situation where the number of IVs exceeds the number of causal parameters of interest. An overidentification test, such as the Sargan test[5] and Hansen's J test,[6] inspect the degree of agreement between different estimators of the causal effect. Although there are several versions of these tests, Newey[7] showed that tests based on a finite set of moment conditions are asymptotically equivalent. The main drawback of these tests is that we do not know which one, if any, of the IVs is valid. Therefore, we cannot determine which of the estimators converge to the true causal effect. As a result, properties that allow for a valid IV are relegated to assumptions that typically rely on subject-matter knowledge. For example, in the well-known study in political economy, Acemoglu et al[8] estimated the causal effect of income on the level of democracy using past savings as an IV. Acemoglu et al[8] acknowledged that the exclusion assumption might be compromised since the saving rates might be correlated with anticipated regime change. Furthermore, an exogenous factor may influence both the saving rates and the democracy level, violating the exogeneity assumption. Another example is epidemiological studies that use the Mendelian randomization approach to estimate causal effects. This approach uses a genetic marker that is associated with a particular exposure and affects an outcome of interest only through the exposure. For instance, VanderWeele et al[9] present a study of the causal effect of smoking on lung cancer by using genetic variants on chromosome 15 as an IV that affects the number of smoked cigarettes per day. However, this genetic variant also affects directly the probability of lung cancer and therefore violates the exclusion assumption.[10]

If any of the IV assumptions are violated, the causal effect cannot be fully identified. Therefore, assessing the robustness of the obtained results to violations of valid IV assumptions is desirable. This assessment can be done, for example, with a sensitivity analysis. The need for sensitivity analysis to violations of the IV assumptions has been addressed in several previous publications.[6,9,11,12] For example, Angrist et al[11] provide rigorous analysis of the IV estimand when the exclusion assumption is violated. Wang et al[13] introduced a method for sensitivity analysis that is based on the Anderson-Rubin[14] (AR) test. The AR test is a uniformly most powerful test used to test the null hypothesis of no causal effect of the exposure on the outcome. Wang et al[13] considered two sensitivity parameters; one for the exclusion and another for the exogeneity assumption. By assuming linearity of the functional form of both violations, one can add up the two parameters and construct a single parameter that captures both the exclusion and the exogeneity assumptions violations. Canley et al[15] also considered one sensitivity parameter that is assumed to account for the two types of violations. The sensitivity analysis of the estimators was performed for a linear model by assuming a different set of values or a prior distribution of the sensitivity parameter. Imbens[16] performed a sensitivity analysis of the exogeneity assumption violation using two distinct sensitivity parameters in the framework of a linear causal model. Additional research on sensitivity analysis was conducted by Imbens & Rosenbaum,[17] Small & Rosenbaum[18] and others.[13,19] However, all these studies dealt only with least-squares-based estimators in linear causal models.[6,11-13,16]

This article starts with formulating valid IV assumptions using Rubin's potential outcomes framework.[11,20] Then, we formulate violations of these assumptions using a single sensitivity parameter $\alpha$ that captures violations of both the exclusion and the exogeneity assumptions. Using a single parameter requires specifications of fewer parameters in the model which is considered a valuable advantage. Next, we conduct a simulation study to illustrate the sensitivity analysis in linear and logistic causal models. Finally, we provide a real-world data example that illustrates the new method.

## 2 | NOTATIONS, DEFINITIONS AND ASSUMPTIONS

### 2.1 | The structural mean model

Let $Y_x$ be the potential outcome[20] of $Y$ when the exposure $X$ is set to $x$. In addition, let $L$ be a set of measured confounders for $X$ and $Y$, $Z$ is the instrumental variable, and $\psi$ is a vector of the causal parameters. See Figure 1 for a graphical illustration of a causal model with a valid IV. A structural mean model[21,22] that parameterizes the average causal effect

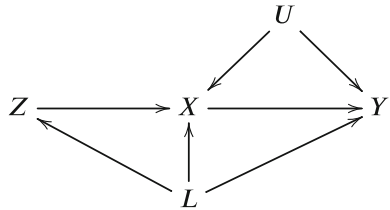

**FIGURE 1**    A causal structure of a valid IV. $Y$ is the outcome variable, $X$ is exposure, and $Z$ is the instrument. $U$ represents all unmeasured confounders of $X$ and $Y$, whereas $L$ represents all measured confounders. The instrument $Z$ affects $Y$ only through the exposure $X$, and is not confounded with the outcome by the unmeasured variables.

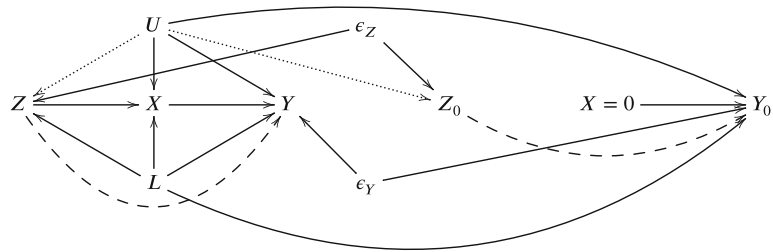

**FIGURE 2**    A twin causal network with an invalid IV. The solid arrows represent the underlying causal structure without the violations. $Y$ is the outcome variable, $X$ is exposure, and $Z$ is the instrument. $U$ represents all unmeasured confounders of $X$ and $Y$, whereas $L$ represents all measured confounders. The dashed arrows from $Z$ to $Y$ and from $Z_0$ to $Y_0$ represent the exclusion assumption violation in the factual and hypothetical networks, respectively. The dotted arrows from $U$ to $Z$ and $Z_0$ represent the exogeneity assumption violation in the factual and the hypothetical networks, respectively. Notably, the factual instrument equals the potential instrument, that is, $Z = Z_0$, since it is determined in both networks prior to the treatment $X$ by the same unmeasured confounder $U$ and the noise term $\epsilon_Z$ (given the measured confounders $L$). The noise terms $\epsilon_Z$ and $\epsilon_Y$ represent all other factors that determine the value of the instrument $Z$, the outcome $Y$, and the potential outcome $Y_0$.

for individuals exposed to level $x$ of $X$, is defined as

$$\xi(E[Y_x|L, Z, X = x]) - \xi(E[Y_0|L, Z, X = x]) = m^T(L)x\psi, \tag{1}$$

where $\xi$ is a link function of a generalized linear model, and $\dim(m(L)) = \dim(\psi)$. Notably, $E[Y_x|L, Z, X = x] = E[Y|L, Z, X = x]$, therefore this part of the model is identifiable from the observed data, whereas identification of the counterfactual mean $E[Y_0|L, Z, X = x]$ requires a valid IV. The composition of the vector-valued function $m(L)$ defines the exact form of the causal model and may allow for interactions between $L$ and the exposure $X$. For example, in a linear model, $\xi$ is the identity link function. Furthermore, assuming $m(L) = 1$ and a binary exposure $X$, simplifies the structural mean model

$$E[Y_1|L, Z, X = 1] - E[Y_0|L, Z, X = 1] = \psi. \tag{2}$$

In such a case, $\psi$ is the average causal effect of the exposure for the exposed. In this scenario, $\psi$ can be consistently estimated using the two-stage least squares (TSLS) method.[23]

We use the twin network to demonstrate the counterfactual implications of violations of the IV assumptions. A twin network is a graphical method that presents two networks together-one for the hypothetical world where the exposure is set to a fixed level $x$ for everyone, and the other for the factual world where the exposure varies randomly. A directed acyclic graph (DAG) of a twin network in Figure 2 provides a graphical illustration of a causal model with unmeasured confounders $U$ and an instrumental variable $Z$. The solid arrows represent the underlying causal structure without the violations. In other words, if we keep only the solid arrows in the DAG, we remain within a network where the IV $Z$ is valid. The dashed arrows from $Z$ to $Y$ and from $Z_0$ to $Y_0$ represent the exclusion assumption violation in the factual and the hypothetical networks, respectively. Namely, the dashed arrows represent the direct effect of the instrument on the outcome in the factual and hypothetical networks, respectively. Notably, the factual instrument equals the potential instrument, that is, $Z = Z_0$, since it is determined in both networks prior to the treatment $X$ by the same unmeasured confounder $U$ and the noise term $\epsilon_Z$ (given the measured confounders $L$). The dotted arrows from $U$ to $Z$ and $Z_0$ represent the exogeneity assumption violation in the factual and the hypothetical networks, respectively. In other words, if the instrument $Z$ ($Z_0$) is associated with the unmeasured confounders $U$ that also confound the exposure $X$ and the outcome $Y$, then the instrument becomes an endogenous variable. It is worth mentioning that the noise terms $\epsilon_Z$ and $\epsilon_Y$ represent all other factors that determine the value of the instrument $Z$, the outcome $Y$, and the potential outcome $Y_0$. Given $\epsilon_Z$, $L$ and $U$, the value of the instrument $Z$ is set deterministically. However, since we cannot measure every possible parent variable, we regard $Z$ as a random variable. Namely, even if the values of $U$ are known, $Z$ remains a random variable where $\epsilon_Z$ accounts for the unexplained variation of $Z$. The same explanation holds for $Y$ ($Y_0$), $U$ and $\epsilon_Y$. Another important feature of the twin network method is that it provides a graphical way for testing independence among counterfactual quantities. According to Figure 2, a counterfactual implication of exogeneity and exclusion assumptions is that a valid IV $Z$ satisfies $Y_0 \perp Z|L$.[24] In other words, if $Z$ is a valid IV, there is no open path between the potential outcome $Y_0$ and the instrument $Z$, conditionally on $L$.[25] An important advantage of this formulation is that it avoids explicit reference to unmeasured confounder $U$, allowing for greater generality.

## 2.2 | G-estimation

Loosely speaking, the G-estimator[26] is a value of the causal parameter $\psi$ under which the assumption $Y_0 \perp Z|L$ holds in the observed data. This value is obtained as a solution to a system of estimating equations.[27] In heterogeneous populations, the value of $Y_0$ is determined by the causal effects specified in the model and the noise term $\epsilon_Y$ that incorporates the set of all other exogenous unknown factors that determine the outcome. This factor is regarded as a disturbance and thus is suppressed in the definition of the causal model. The potential outcome $Y_0$ is counterfactual, namely, an unobserved quantity for the exposed individuals. Thus, we can only estimate its mean value. For example, in a linear model, if the causal effect for the exposed is defined as $m^T(L)x\psi$, then subtracting it from the observed outcome $Y$, results, on average, in the potential outcome $Y_0$. We denote this quantity by $h(\psi)$, where the exact form of $h(\psi)$ depends on the link function $\xi$, such that

$$h(\psi) = \begin{cases} Y - m^T(L)x\psi, & \text{if } \xi \text{ is the identity link function} \\ Y \exp\{-m^T(L)x\psi\}, & \text{if } \xi \text{ is the log link function} \\ \text{expit}\big(\text{logit}(E[Y|X=x,Z,L] - m^T(L)x\psi\big), & \text{if } \xi \text{ is the logit link function.} \end{cases} \tag{3}$$

G-estimators are obtained in the same manner in generalized linear models, following the transformation that is defined as the inverse of the link function. Formally, the G-estimator solves the estimating equation

$$\sum_{i=1}^{n} D(L_i, Z_i)h_i(\psi) = 0, \tag{4}$$

where $D(L, Z)$ is an arbitrary function with the same dimension as $\psi$, such that $E[D(L, Z)|L] = 0$. Two common choices for $D$ are[22]

$$D(L, Z) = m(L)(Z - E[Z|L]) \tag{5}$$

and

$$D(L, Z) = m(L)(E[X|L, Z] - E[X|L]). \tag{6}$$

The function in Equation (5) can only be used for one-dimensional instruments, while the function in Equation (6) can be used for multidimensional IVs as well. However, this advantage comes with a price—it requires specification of an additional model. Particularly, the function in Equation (5) requires a model only for $E[Z|L]$, whereas the function in Equation (6) requires a model for both $E[X|L, Z]$ and $E[X|L]$. Without loss of generality, from now on, we will assume a one-dimensional IV and therefore consider the more simple function as in Equation (5). Consistency of the G-estimator relies on the validity of the IV, namely, on the conditional independence of $h(\psi)$ and $Z$, given $L$. A possible representation of a less strict condition of mean independence between $h(\psi)$ and $Z$ is via the following equation[28,29]

$$E[h(\psi)|L, Z] = E[h(\psi)|L] = a(L). \tag{7}$$

If the conditional mean independence does not hold, then Equation (7) does not hold either. Although a valid IV satisfies complete independence of $Y_0$ and $Z$, violations that affect, for instance, only the variance of the variables will not affect the estimators' consistency of the structural mean model parameters. To avoid confusion, by 'violation' we refer to settings that violate the mean independence of $Y_0$ and $Z$, given $L$. The conditional dependence of $Y_0$ and $Z$ can arise from the violation of the exclusion assumption, the violation of the exogeneity assumption, or both. To illustrate this claim, we use the twin network in Figure 2. For example, if the exclusion assumption is violated, it implies a direct effect of the IV $Z$ on the outcome $Y$. Since the potential IV $Z_0$ possess the same parent variable $\epsilon_Z$ (in addition to the measured confounders $L$), this violation implies an association between the potential outcome $Y_0$ and the observed instrument $Z$. If the exogeneity assumption is violated, it means that both $Z$ and $Y_0$ have a common parent $U$. If $U$ is unmeasured and thus not controlled, it creates an association between $Z$ and $Y_0$. A possible consequence of these violations is inconsistency of the G-estimator of $\psi$.

## 2.3 | Sensitivity analysis with a single parameter $\alpha$

We model compositions of mean independence violations using a parametric function $b(L, Z; \alpha)$, where for a generalized linear model, one can use the following form

$$\xi(E[Y_0|L, Z]) = g(L, Z) = a(L) + b(L, Z; \alpha). \tag{8}$$

Since any source of violation will result in an association between $Y_0$ and $Z$ (given $L$), we can use a scalar function $b(L, Z; \alpha)$, where $\alpha$ is a parameter that incorporates this association. Therefore, we require that $b(L, Z; 0) = 0$. Namely, the violation is eliminated if there is no association between $Y_0$ and $Z$. Consequently, $b(L, Z; \alpha)$ is non-zero for $\alpha \neq 0$. Defining a sensitivity parameter in terms of the deviation from a counterfactual independence assumption is standard in the causal inference literature, and used, for instance, in sensitivity analysis for truncation by death,[30] marginal structural models[31] and transportability in randomized trials.[32] Since $|b(L, Z; \alpha)|$ expresses the magnitude of the assumptions' violation, we can expect that for given values of $L$ and $Z$, $|b(L, Z; \alpha)|$ will be a monotonically non-decreasing function of $|\alpha|$. For example, in a linear violation structure such that $b(L, Z; \alpha) = Z\alpha$, $\alpha$ incorporates the association's strength and direction. For readers who prefer to define interpret the sensitivity parameter in terms of causal mechanisms (eg, arrows on a DAG), we show in the Appendix how our parameter $\alpha$ can be decomposed into a direct effect of the instrument on the outcome, and an association between the instrument and the outcome due to confounding. For linear models, Wang et al[13] considered such a parametrization, and used this for sensitivity analysis. Our results in the Appendix thus extend those by Wang et al. and allow the reader to specify $\alpha$ directly or indirectly in terms of parameters similar to those proposed by Wang et al.

The interpretation of the direct effect is conceptually straightforward. However, the confounded association is more subtle since assessing the strength (absolute value) and direction (sign) of $\alpha$ induced by omitted confounders can be tricky. In order to give meaningful values or even crude bounds on $\alpha$, subject-matter knowledge is required. For the aforementioned linear structure, a subject-matter expert can answer two main questions: (1) Is a direct causal effect of the instrument on the outcome plausible? (If the answer is affirmative—how strong is the effect, and what is its direction?) (2) Are there any important confounders that can be measured and therefore mitigate the possible bias caused by their omission? For example, in a special case of an RCT with imperfect compliance, one can rule out the exclusion assumption violation by the very design of the experiment. In addition, as a well-conducted experiment strives to account for every known confounder, one can assume that even if a violation occurs, it can only be due to the omission of non-essential confounders (exogeneity assumption violation) and, therefore, should not be severe. In such a case in the sensitivity analysis, one may consider only small values of $\alpha$ around 0.

Using Equation (3) in the estimating equations for $\alpha \neq 0$ will result in an inconsistent estimator since the estimating equations in such a setting will be biased w.r.t. 0. To resolve this pitfall, we reformulate $h(\psi)$ from Equation (3) as a function of the parameter $\alpha$

$$h(\psi;\alpha) = \begin{cases} Y - m^T(L)x\psi - b(L,Z;\alpha), \\ Y \exp\{-m^T(L)x\psi - b(L,Z;\alpha)\}, \\ \text{expit}\big(\text{logit}(E[Y|X=x,Z,L] - m^T(L)x\psi - b(L,Z;\alpha)\big), \end{cases} \tag{9}$$

such that $E[h(\psi;\alpha)|Z,L] = E[h(\psi;\alpha)|L] = a(L)$, for the true value of $\alpha$ that is denoted by $\alpha = \alpha^*$ (see proof of Lemma 1 in Appendix). In other words, for $\alpha^*$, the assumption $Y_0 \perp Z|L$ is still violated; however, the estimating functions remain unbiased, and thus the G-estimator is consistent w.r.t. the true causal parameter $\psi$. Since the consistency depends on the correct specification of $\alpha$ in $h(\psi;\alpha)$ and $b(L,Z;\alpha)$, we consider $\alpha$ as a sensitivity parameter. The exact value of $\alpha^*$ would typically not be known to the analyst. Therefore, a sensitivity analysis can be carried out by varying $\alpha$ over a range of values considered plausible by subject-matter experts, and estimating the causal exposure effect $\psi$ separately for each value of $\alpha$.

## 2.4 | Asymptotic variance and distribution of the G-estimator

Let $\sum_{i=1}^n Q(Y_i,X_i,L_i,Z_i;\theta,\alpha) = 0$ be the estimating equations of a parametric causal model, where the vector of estimands is $\theta^T = (\beta_Y^T, \mu_Z, \psi)$, and

$$Q(Y,X,L,Z;\theta,\alpha) = \begin{pmatrix} S(Y,X,L;\beta_Y) \\ S(L,Z;\mu_Z) \\ D(L,Z;\mu_Z)h(\psi;\alpha) \end{pmatrix}. \tag{10}$$

Notably, $S(Y,X,L;\beta_Y)$ are unbiased estimating functions of $\beta_Y$. Namely, $E[S(Y,X,L;\beta_Y)] = 0$ for the true value of $\beta_Y$. The model $E[Y|X,L,Z;\beta_Y]$ for the outcome $Y$ is required when $\xi$ is the logit link function. In such a case, the function $h(\psi;\alpha)$ (9) in the estimating equations depends on the outcome model. Therefore, $S(Y,X,L;\beta_Y)$ is the score function of the logistic model that is used to estimate the parameters $\beta_Y$. In contrast to the logistic regression model, for the identity link function, no outcome model is needed in $h(\psi;\alpha)$ (9), hence we do not have the unbiased estimating functions $S(Y,X,L;\beta_Y)$ in the estimating equations $Q(Y,X,L,Z;\theta,\alpha)$. In addition, the function $S(L,Z;\mu_Z)$ is an unbiased estimating function of the instrument model $E[Z|L;\mu_Z]$, and $D(L,Z;\mu_Z)$ is as defined in Equation (5).

The asymptotic variance of $\hat{\theta}$ is given by the sandwich formula[27]

$$V(\theta_0,\alpha) = n^{-1}A(\theta_0,\alpha)^{-1}B(\theta_0,\alpha)A(\theta_0,\alpha)^{-T} \tag{11}$$

where $A(\theta_0,\alpha) = E[-\partial Q(\theta_0,\alpha)/\partial\theta^T]$, $B(\theta_0,\alpha) = E[Q(\theta_0,\alpha)Q(\theta_0,\alpha)^T]$, and $\theta_0$ is the true value of the unknown parameters. It can be shown that the upper-left block of the "meat" matrix $B(\theta_0,\alpha)$ is the expected value of the outer product of the unbiased estimating functions

$$E[S(Y,X,L;\beta_Y)S(Y,X,L;\beta_Y)^T].$$

For example, for one-dimensional $\mu_Z$ and $\psi$, and linear $E[Z|L;\mu_Z]$ such that $S(L,Z;\mu_Z) = D(L,Z;\mu_Z)$, the lower right $2\times2$ block of $B(\theta_0,\alpha)$ is

$$E\begin{pmatrix} D^2(L,Z;\mu_Z) & D^2(L,Z;\mu_Z)h(\psi;\alpha) \\ D^2(L,Z;\mu_Z)h(\psi;\alpha) & D^2(L,Z;\mu_Z)h^2(\psi;\alpha) \end{pmatrix}. \tag{12}$$

The vector of $\theta_0$ estimators is denoted by $\hat{\theta}(\alpha)$. This asymptotic covariance results in a robust estimator of the coefficients' variance w.r.t. model misspecification. The asymptotic distribution of the estimators $\hat{\theta}(\alpha)$ is multivariate normal,[27] namely

$$\sqrt{n}(\hat{\theta}(\alpha) - \theta_0(\alpha)) \xrightarrow{D} \mathcal{N}_p(0, V(\theta_0,\alpha)), \tag{13}$$

where the subscript $p$ denotes the dimension of the parametric space $\theta \in \Theta$, and the superscript $D$ denotes convergence in distribution. For the sample version of "bread" and "meat" matrices, we replace the expectation operator with the corresponding sample means, the true $\beta_0^T = (\beta_Y^T, \mu_Z)$ with the corresponding unbiased estimators, and $\psi$ with its G-estimator that is obtained for a given value of $\alpha$.

## 2.5 | Examples

### 2.5.1 | Linear causal model

In order to illustrate the implication of nonzero value $\alpha^*$, we start with a linear model with explicit unmeasured confounders $U$, and possible violations of the exclusion and the exogeneity assumptions. Namely, we assume an underlying causal structure as illustrated in the DAG in Figure 2. However, for clarity of exposition, we assume no measured confounders $L$, that is, $L = \emptyset$. It can be shown (see, eg, section 2 Vansteelandt & Didelez[24]) that the structural mean model in Equation (2) can be obtained by averaging out the following structural mean model

$$E[Y_x|Z, U, X = x] - E[Y_0|Z, U, X = x] = x\psi \tag{14}$$

w.r.t. the conditional distribution of $U$, given $Z$ and $X$. We specify a violation of valid IV assumptions with $b(Z, L; \alpha^*) = \alpha^* Z$. This specification may represent a direct effect of $Z$ on $Y$, a confounding effect of unmeasured variables $U$ on $Z$ and $Y$, or both. Since $L = \emptyset$, we obtain $D(L, Z; \mu_Z) = S(L, Z; \mu_Z) = Z - \mu_Z$, where $E[Z] = \mu_Z$, and the unbiased estimating function of the G-estimator is

$$D(L, Z; \mu_Z)h(\psi; \alpha^*) = (Z - \mu_Z)(Y - \psi X - \alpha^* Z). \tag{15}$$

Using simple algebra as illustrated in Lemma 2 in the Appendix, the true causal effect can be formulated as a function of the sensitivity parameter $\alpha^*$ in the following way

$$\psi = \frac{\beta_{YZ}}{\beta_{XZ}} - \frac{\alpha^*}{\beta_{XZ}}, \tag{16}$$

where $\beta_{XZ} = cov(X, Z)/var(Z)$, and $\beta_{YZ} = cov(Y, Z)/var(Z)$. For a special case where $U = \emptyset$, the sensitivity parameter $\alpha^*$ is simply the true regression coefficient $\beta_{YZ.X} = cov(Y, Z|X)/var(Z|X)$. Hence, a G-estimator that can be corrected using Equation (16) to remain consistent, coincides with the least squares estimator of $\psi$ (for further details, please refer to Lemma 4 in the Appendix). Following the same logic, in the presence of additive unmeasured confounders, with an underlying causal structure as in Figure 2, the sensitivity parameter $\alpha^*$ is $\beta_{YZ.XU} = cov(Y, Z|X, U)/var(Z|X, U)$. Namely, $\alpha^*$ depends on the correlation of the outcome with the IV conditional on unmeasured confounders $U$ and exposure $X$, and therefore is not point-identifiable. Notably, from Equation (16), we can obtain bounds of the corresponding causal effect by setting bounds on the true value of the sensitivity parameter $\alpha^*$.

**Theorem 1.** *Assume the causal structure as in Figure 1, and a structural mean model as in Equation (2). The true sensitivity parameter $\alpha^*$, as defined in (8), is non-identifiable.*

For a detailed proof of Theorem 1 please refer to Appendix. An immediate implication of Theorem 1 is that, in the presence of unmeasured confounders, the causal effect $\psi$ is non-estimable using the G-estimation method. Therefore, it is desirable to perform a sensitivity analysis where $\alpha$ is varied over a set of values considered plausible by a subject-matter expert. Otherwise, if such values cannot be determined, we suggest starting with a range symmetric values around 0. Each value of $\alpha$ from this set is mapped to a corresponding G-estimator, and this procedure thus provides bounds on the true causal effect.

As illustrated in Equation (13), the asymptotic distribution of the G-estimator is normal, where its asymptotic variance can be computed using the sandwich formula as in Equation (11). For linear outcome models, the "bread" matrix $A(\theta_0, \alpha)$ is block-diagonal matrix since the partial derivatives of $D(L, Z; \mu_Z)$ and $h(\psi; \alpha)$ do not depend on $\beta_Y$. This result simplifies the calculations of the G-estimator variance since the upper-right block derived for the estimators of $\beta_Y$ does not contribute to the variance of the G-estimator. For example, for one-dimensional $\mu_Z$ and $\psi$, and linear $E[Z|L; \mu_Z]$ such that $S(L.Z; \mu_Z) = D(L, Z; \mu_Z)$, the form of the right-lower block of the "bread" matrix $A(\theta_0, \alpha)$ is

$$E\begin{pmatrix} 1 & 0 \\ h(\psi; \alpha) & XD(L, Z; \mu_Z) \end{pmatrix}.$$

## 2.5.2 | Logistic causal model

Assume a binary outcome $Y$, a binary exposure $X$, a binary instrument $Z$, and an unmeasured confounder $U$. In addition, assume that there are no measured confounders, that is, $L = \emptyset$, and $m(L) = 1$. We define the structural mean model on the logit scale

$$\text{logit} P(Y_1 = 1 | X = 1, Z) - \text{logit} P(Y_0 = 1 | X = 1, Z) = \psi. \tag{17}$$

Therefore, we define the violation on the logit scale as well

$$b(Z, L; \alpha^*) = \text{logit} P(Y_0 = 1 | Z) - \text{logit} P(Y_0 = 1 | Z = 0) = \alpha^* Z. \tag{18}$$

Assuming logistic outcome model for $E[Y|Z, X; \beta_Y]$ with an interaction term, such that $\beta_Y^T = (\beta_0, \beta_x, \beta_z, \beta_{xz})$, $S(Y, X; \beta_Y)$ are the score functions of this model, where

$$S(Y, X; \beta_Y) = \begin{pmatrix} (Y - E[Y|Z, X; \beta_Y]) \\ (Y - E[Y|Z, X; \beta_Y])X \\ (Y - E[Y|Z, X; \beta_Y])Z \\ (Y - E[Y|Z, X; \beta_Y])XZ \end{pmatrix}.$$

In addition, let $D(L, Z; \mu_Z) = S(L, Z; \mu_Z) = Z - \mu_Z$. Therefore, the last function $D(L, Z; \mu_Z)h(\psi; \alpha)$ in the system of estimating equations (10) is

$$(Z - \mu_Z)\text{expit}\left(\text{logit} P(Y = 1 | Z, X; \beta_Y) - X\psi - \alpha Z\right). \tag{19}$$

Since $h(\psi; \alpha)$ in the logistic causal model depends on the outcome model $E[Y|Z, X; \beta_Y]$, the variance of the G-estimator is now affected by the variance of $\beta_Y$ estimators. In particular, the "bread" matrix $A(\theta_0, \alpha)$ is no longer a block diagonal as in the linear model, since the partial derivatives of $D(L, Z; \mu_Z)h(\psi; \alpha)$ w.r.t. $\beta_Y$ are non-zero valued, that is,

$$E\left[\frac{\partial}{\partial \theta^T} Dh(\psi; \alpha)\right] = E(Dv_{\psi,\alpha}, XDv_{\psi,\alpha}, ZDv_{\psi,\alpha}, XZDv_{\psi,\alpha}, -h(\psi; \alpha), -XDv_{\psi,\alpha})^T,$$

where $D$ is a shorthand for $D(L, Z; \beta_Y)$, $v_{\psi,\alpha} = h(\psi; \alpha)(1 - h(\psi; \alpha))$, and $\theta^T = (\beta_0, \beta_x, \beta_z, \beta_{xz}, \mu_Z, \psi)$. Therefore, the computations of the covariance matrix are more involved since they require the inversion of high-order matrices. The simulation study in Section 3 illustrates a sensitivity analysis of the G-estimators of causal parameters using an invalid instrument in linear and logistic causal models.

## 3 | SIMULATION STUDY

In many practical situations, we cannot rule out the presence of unmeasured confounders that affect both $X$ and $Y$, and an absence of violation of the IV assumptions, as illustrated in Figure 2. In situations with invalid IVs, the G-estimators of the causal parameter $\psi$ will be asymptotically biased,[12] while their correction is not possible since the true $\alpha^*$ is not point-identifiable. Thus, sensitivity analysis is desirable. To perform the sensitivity analysis, we vary $\alpha$ over a range of plausible values and map them into a range of corresponding G-estimators of $\psi$. Since any violation of valid IV assumptions results in an association between the instrument $Z$ and the potential outcome $Y_0$, we can simulate the appropriate data without explicitly using unmeasured confounders. In other words, instead of using the explicit unmeasured confounders $U$, we start with the structural mean model and the consequence of the violations. Therefore, in the data generating process (DGP), the unmeasured confounders are implicit, whereas the consequence of their omission is explicitly specified. For the linear causal models, we consider a sample size of $n = 1000$, two values of $\psi \in \{0, 1.5\}$, two values of $\alpha^* \in \{0, 0.5\}$, and all their combinations. For the logistic causal models, we consider a sample size of $n = 1000$, two values of $\psi \in \{0, 0.5\}$, two values of $\alpha^* \in \{0, 0.5\}$, and all their combinations. For every model, we solve the estimating equations, as defined in Equation (10), for every $\alpha$ in $\{\alpha^* - c(1 - k/10)\}_{k=0}^{20}$, for $\alpha^* \in \{0, 0.5\}$, and $c = 0.2$. Namely, we map every $\alpha$ to a G-estimator $\hat{\psi}_G(\alpha)$, that is obtained by numerical solution of Equation (15). Furthermore, for every G-estimator $\hat{\psi}_G(\alpha)$, we compute a 95%-level asymptotically-correct CI using the sandwich formula as in Equation (11). We repeat each simulation $m = 1000$ times, and compute the empirical coverage rates and the mean length of the 95%-level CIs for the true causal parameter $\psi$ for every $\alpha$ in the aforementioned sequence. The coverage rates are visually illustrated in Figures 3 and 6 for every combination of $\psi$ and $\alpha$. The distribution of the 95%-level CIs are illustrated in Figures 5 and 8, and Tables 1 and 2. Additionally, to illustrate the bias of $\hat{\psi}_G(\alpha)$ as a function of $\alpha$, we present in Figures 4 and 7 boxplots of the empirical distribution of $\hat{\psi}_G(\alpha)$ for every $\alpha$. Further model-related specifications are presented in the following subsections. A summary of the simulation study and discussion appear at the end of this section. The simulations source code is available on the authors' GitHub repository.[†]

## 3.1 | Linear causal model

For the linear causal model simulation, we assume a normally distributed outcome $Y$, a binary exposure $X$, and a binary instrument $Z$. Assume that the structural mean model is as specified in Equation (14), and the violation structure is as defined in Equation (8). The DGP is set as follows

$$
\begin{aligned}
Z &\sim Ber(p_z) \\
X|Z = z &\sim Ber(\text{expit}(\gamma_0 + \gamma_z z)) \\
Y|X = x, Z = z &\sim N(\beta_0 + \beta_x x + \beta_z z + \beta_{xz} xz, \sigma^2).
\end{aligned}
\tag{20}
$$

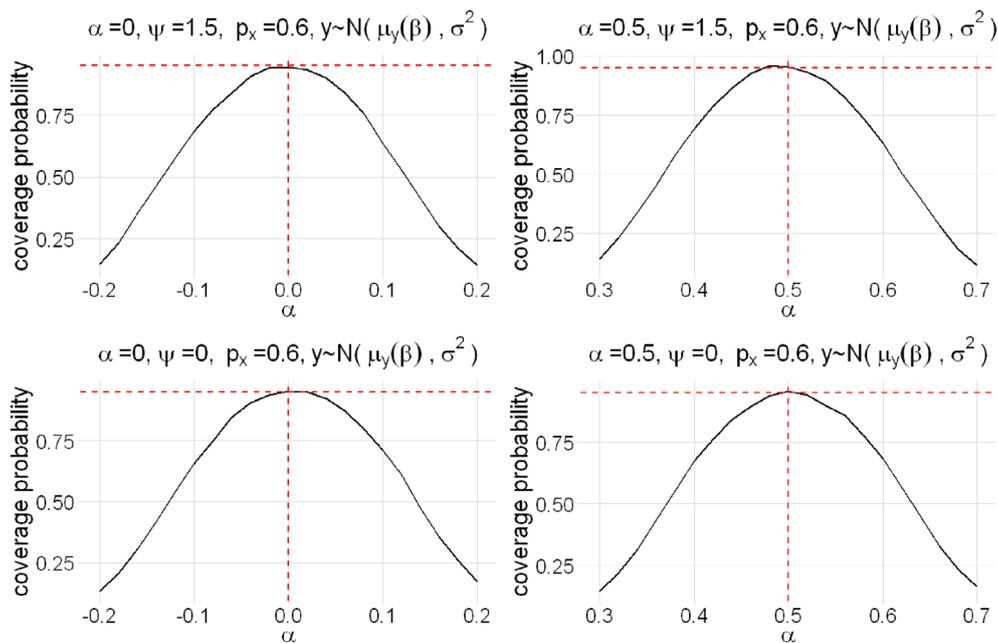

**FIGURE 3** Coverage rates of the 95%-level CIs of the true causal parameter $\psi \in \{0, 1.5\}$ in the linear causal model as a function of the sensitivity parameter $\alpha$, for $\alpha^* \in \{0, 0.5\}$, for a sample size $n = 1000$, and $m = 1000$ repetitions. The linear outcome model is $Y|X = x, Z = z \sim N(\mu_Y(\beta), \sigma^2)$, where $\mu_Y(\beta) = \beta_0 + \beta_x x + \beta_z z + \beta_{xz} xz$ as in (20), and $p_x = P(X = 1) = 0.6$. The dashed red horizontal line denotes the 95% coverage rate. The dashed vertical red line denotes the true value of $\alpha$. The solid black line denotes the computed coverage probabilities.

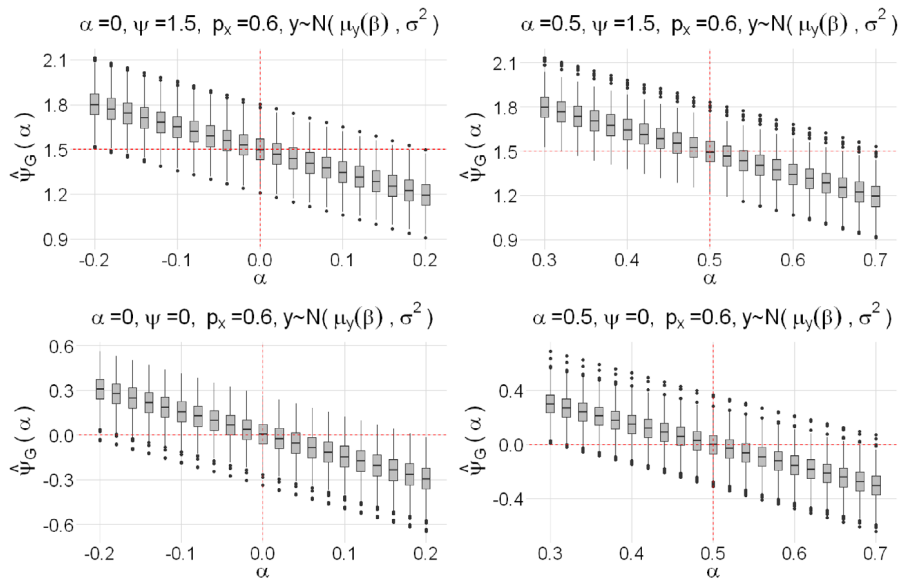

**FIGURE 4** Boxplots of empirical distribution of the G-estimators $\hat{\psi}_G(\alpha)$ of the true causal parameter $\psi \in \{0, 1.5\}$ in a linear causal model as a function of the sensitivity parameter $\alpha$, for $\alpha^* \in \{0, 0.5\}$, for sample size $n = 1000$, and $m = 100$ repetitions. The linear outcome model is $Y|X = x, Z = z \sim N(\mu_Y(\beta), \sigma^2)$, where $\mu_Y(\beta) = \beta_0 + \beta_x x + \beta_z z + \beta_{xz} xz$ as in (20), and $p_x = P(X = 1) = 0.6$. The dashed horizontal red line denotes the true value of $\psi$. The dashed vertical red line denotes the true value of $\alpha$.

**TABLE 1** Mean values of the 95%-level CIs lengths for $\psi = 0, 1.5$ as a function of $\alpha$ in the linear causal model: The true $\alpha^* = 0, 0.5$, respectively. $p_x = 0.6$.

| $\alpha$ | $\psi = 0$ | $\psi = 1.5$ | $\alpha$ | $\psi = 0$ | $\psi = 1.5$ |
|---|---|---|---|---|---|
| −0.2 | 0.391 | 0.4 | 0.3 | 0.386 | 0.392 |
| −0.18 | 0.392 | 0.399 | 0.32 | 0.386 | 0.391 |
| −0.16 | 0.392 | 0.398 | 0.34 | 0.387 | 0.39 |
| −0.14 | 0.393 | 0.396 | 0.36 | 0.388 | 0.389 |
| −0.12 | 0.393 | 0.395 | 0.38 | 0.388 | 0.388 |
| −0.1 | 0.394 | 0.394 | 0.4 | 0.389 | 0.387 |
| −0.08 | 0.394 | 0.393 | 0.42 | 0.39 | 0.386 |
| −0.06 | 0.395 | 0.392 | 0.44 | 0.391 | 0.385 |
| −0.04 | 0.396 | 0.391 | 0.46 | 0.392 | 0.384 |
| −0.02 | 0.396 | 0.39 | 0.48 | 0.393 | 0.383 |
| 0 | 0.397 | 0.389 | 0.5 | 0.394 | 0.382 |
| 0.02 | 0.398 | 0.388 | 0.52 | 0.395 | 0.382 |
| 0.04 | 0.399 | 0.387 | 0.54 | 0.396 | 0.381 |
| 0.06 | 0.4 | 0.386 | 0.56 | 0.397 | 0.38 |
| 0.08 | 0.401 | 0.385 | 0.58 | 0.398 | 0.38 |
| 0.1 | 0.402 | 0.384 | 0.6 | 0.399 | 0.379 |
| 0.12 | 0.403 | 0.384 | 0.62 | 0.4 | 0.378 |
| 0.14 | 0.404 | 0.383 | 0.64 | 0.402 | 0.378 |
| 0.16 | 0.405 | 0.382 | 0.66 | 0.403 | 0.378 |
| 0.18 | 0.406 | 0.382 | 0.68 | 0.404 | 0.377 |
| 0.2 | 0.407 | 0.381 | 0.7 | 0.405 | 0.377 |

**TABLE 2** Mean values of the 95%-level CIs lengths for $\psi = 0, 0.5$ as a function of $\alpha$ in the logistic causal model: The true $\alpha^* = 0, 0.5$, respectively. $p_y = 0.3$, and $p_x = 0.6$.

| $\alpha$ | $\psi = 0$ | $\psi = 0.5$ | $\alpha$ | $\psi = 0$ | $\psi = 0.5$ |
|---|---|---|---|---|---|
| −0.2 | 0.86 | 0.972 | 0.3 | 0.912 | 1.043 |
| −0.18 | 0.857 | 0.967 | 0.32 | 0.908 | 1.039 |
| −0.16 | 0.853 | 0.963 | 0.34 | 0.905 | 1.035 |
| −0.14 | 0.849 | 0.958 | 0.36 | 0.901 | 1.031 |
| −0.12 | 0.846 | 0.953 | 0.38 | 0.898 | 1.027 |
| −0.1 | 0.843 | 0.949 | 0.4 | 0.895 | 1.023 |
| −0.08 | 0.839 | 0.944 | 0.42 | 0.891 | 1.019 |
| −0.06 | 0.836 | 0.94 | 0.44 | 0.888 | 1.015 |
| −0.04 | 0.833 | 0.936 | 0.46 | 0.885 | 1.01 |
| −0.02 | 0.83 | 0.931 | 0.48 | 0.882 | 1.006 |
| 0 | 0.828 | 0.927 | 0.5 | 0.88 | 1.002 |
| 0.02 | 0.825 | 0.923 | 0.52 | 0.877 | 0.998 |
| 0.04 | 0.822 | 0.918 | 0.54 | 0.874 | 0.994 |
| 0.06 | 0.82 | 0.914 | 0.56 | 0.872 | 0.99 |
| 0.08 | 0.818 | 0.91 | 0.58 | 0.869 | 0.986 |
| 0.1 | 0.815 | 0.906 | 0.6 | 0.867 | 0.982 |
| 0.12 | 0.813 | 0.902 | 0.62 | 0.865 | 0.978 |
| 0.14 | 0.811 | 0.898 | 0.64 | 0.863 | 0.974 |
| 0.16 | 0.809 | 0.895 | 0.66 | 0.861 | 0.97 |
| 0.18 | 0.807 | 0.891 | 0.68 | 0.859 | 0.966 |
| 0.2 | 0.805 | 0.887 | 0.7 | 0.857 | 0.963 |

We specify the marginal distribution of $X$ and $Z$. Thus, five out of the seven parameters of the DGP can be set freely. In particular, we set $P(Z = 1) = p_z = 0.5$, and $P(X = 1) = p_x = 0.6$. In order to relate the violation structure to the DGP parameters of the observed data, we use the fact that a valid IV satisfies $Y_0 \perp Z|L$; particularly, it implies a mean independence $E[Y_0|Z = 1, L] = E[Y_0|Z = 0, L]$. Therefore, for $L = \emptyset$ and an invalid IV with the violation structure $b(L, Z; \alpha^*) = \alpha^* Z$, we obtain

$$E[Y_0|Z = 1] - E[Y_0|Z = 0] = \alpha^*. \tag{21}$$

We use the DGP presented in (20) to express $E[Y_0|Z]$ as a function of the observed data parameters and the causal parameter $\psi$

$$E[Y_0|Z = z] = E[E[Y_0|Z = z, X]|Z = z]$$
$$= (\beta_0 + \beta_x x)(1 - \text{expit}(\gamma_0 + \gamma_z z)) + (\beta_0 + \beta_x x + \beta_z z + \beta_{zx} zx - \psi)\text{expit}(\gamma_0 + \gamma_z z) .$$

Notably, we obtain a function w.r.t. the unknown parameters that characterize the DGP of $Y$. By plugging in this result in Equation (21) for $Z = 0$ and $Z = 1$, we obtain the functional relationship between the sensitivity parameter $\alpha^*$ and the DGP parameters of the observed data. Please refer to Appendix for a detailed derivation. Several of these parameters we set freely, while the others are calculated to satisfy the specified values of the marginal probabilities of $Z = 1$ and $X = 1$. These specifications leave three degrees of freedom for the $\beta_Y^T = (\beta_0, \beta_x, \beta_z, \beta_{zx})$ vector, and one degree of freedom for the $\gamma^T = (\gamma_0, \gamma_z)$ vector. Without loss of generality, we set $\beta_0 = \beta_x = \beta_z = 1$, and solve for $\beta_{zx}$. Furthermore, we set $\gamma_0 = -1$, and solve for $\gamma_z$. Figure 3 illustrates the coverage rates of the 95%-level CI for the true causal parameter $\psi$ as a function of $\alpha$. Figure 4 illustrates the bias of the G-estimators as a function of $\alpha$. Figure 5 illustrates the distribution of the lengths of the 95%-level CIs. Table 1 presents the mean values of 95%-level CIs lengths as a function of $\alpha$.

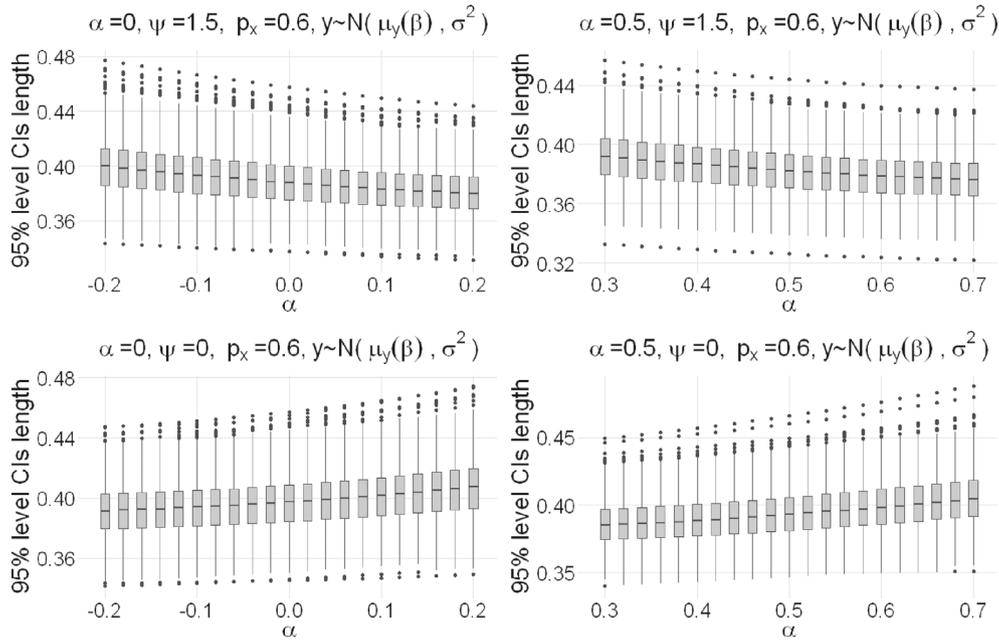

**FIGURE 5** Boxplots of the 95%-level CIs length for $\psi \in \{0, 1.5\}$ in a linear causal model as a function of the sensitivity parameter $\alpha$, for $\alpha^* \in \{0, 0.5\}$, for a sample size $n = 1000$, and $m = 1000$ repetitions. The linear outcome model is $Y|X = x, Z = z \sim N(\mu_Y(\beta), \sigma^2)$, where $\mu_Y(\beta) = \beta_0 + \beta_x x + \beta_z z + \beta_{xz} xz$ as in (20), and $p_x = P(X = 1) = 0.6$.

## 3.2 | Logistic causal model

For the logistic causal model simulation, we consider a binary outcome $Y$, a binary exposure $X$, and a binary instrument $Z$. The structural mean model is as specified in Equation (17), and the violation structure is as specified in Equation (18). The DGP is given below

$$Z \sim Ber(p_z)$$
$$X|Z = z \sim Ber(\text{expit}(\gamma_0 + \gamma_z z))$$
$$Y|X = x, Z = z \sim Ber(\text{expit}(\beta_0 + \beta_x x + \beta_z z + \beta_{xz} xz)). \quad (22)$$

We specify the marginal distribution of $Y$, $X$, and $Z$. In order to relate the violation structure to the DGP parameters of the observed data we use the fact that a valid IV satisfies $Y_0 \perp Z|L$, particularly, $P(Y_0|Z = 1) = P(Y_0|Z = 0)$. Therefore, for an invalid IV with the violation structure as specified in Equation (18), we obtain

$$\text{logit} P(Y_0|Z = 1) - \text{logit} P(Y_0|Z = 0) = \alpha^*. \quad (23)$$

By using the causal parameter $\psi$ and the DGP presented in (22), we obtain a function w.r.t. the unknown parameters of the observed data

$$P(Y_0 = 1|Z = z) = \text{expit}(\beta_0 + \beta_z z)(1 - \text{expit}(\gamma_0 + \gamma_z z)) + \text{expit}(\beta_0 + \beta_x x + \beta_z z + \beta_{xz} xz - \psi)(1 - \text{expit}(\gamma_0 + \gamma_z z)). \quad (24)$$

By plugging in this result in Equation (23), we obtain the functional relationship between the sensitivity parameter $\alpha^*$ and the DGP parameters of the observed data. Please refer to Appendix for a detailed derivation. Several of these parameters we set freely, while the others are calculated to satisfy the specified values of the marginal probabilities of $Z = 1$, $X = 1$, and $Y = 1$. In particular, we set $P(Z = 1) = p_z = 0.5$, and $P(X = 1) = p_x = 0.6$. We consider two marginal probabilities of $p_y \in \{0.3, 0.8\}$. These specifications leave two degrees of freedom for the $\beta_Y^T = (\beta_0, \beta_x, \beta_z, \beta_{xz})$ vector, and one degree of freedom for the $\gamma^T = (\gamma_0, \gamma_z)$ vector. Without loss of generality, we set $\beta_x = \beta_z = 1$, and solve for $\beta_0$ and $\beta_{xz}$. Furthermore, we set $\gamma_0 = -1$, and solve for $\gamma_z$. Figures 6 and A1 illustrate the coverage rates of the 95%-level CIs of the true causal parameter $\psi$ as a function of $\alpha$, for $p_y = 0.3, 0.8$, respectively. Figures 7 and A2 illustrate the bias of the G-estimators as a function of $\alpha$, for $p_y = 0.3, 0.8$, respectively. Figures 8 and A3 illustrate the distribution of the 95%-level CIs length. Tables 2 and A1 present the mean values of 95%-level CIs as a function of $\alpha$. Figures and Tables of the simulation analysis for $p_y = 0.8$ appear in Appendix.

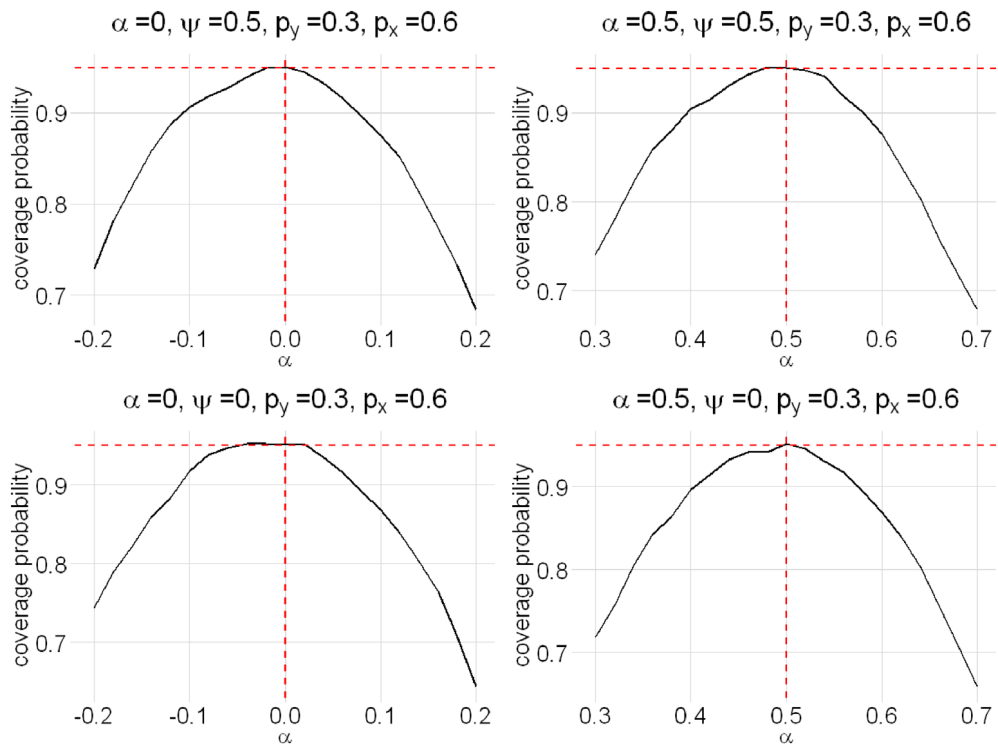

**FIGURE 6** Coverage rates of the 95%-level CIs of the true causal parameter $\psi \in \{0, 0.5\}$ in the logistic causal model as a function of the sensitivity parameter $\alpha$, for $\alpha^* \in \{0, 0.5\}$, for sample size $n = 1000$, and $m = 1000$ repetitions. The logistic outcome model $E[Y|X, Z; \beta_Y]$ is as given in (22), and $P(Y = 1) = p_y = 0.3$. The dashed horizontal red line denotes the 95% coverage rate. The dashed vertical red line denotes the true value of $\alpha$. The solid black line denotes the computed coverage probabilities.

## 3.3 | Simulation summary

In the simulation study, parameters of the DGP are determined after the specification of the true $\psi$ and $\alpha^*$. Therefore, we can use $\alpha^*$ in the estimating equations to construct consistent G-estimators of $\psi$. In real-world applications, $\alpha^*$ is rarely known. Thus, finding a plausible interval for $\alpha$ can be challenging since such an interval depends on subject-matter knowledge.

1. The proposed sensitivity analysis method of G-estimators works well for valid and invalid IVs, and for linear and logistic models. Namely, if $\alpha^* = 0$, then the G-estimator of $\psi$ is consistent for $\alpha = 0$, and its 95%-level CI attains approximately its nominal coverage rate.
2. The proposed sensitivity analysis method of G-estimators works well when the true causal effect $\psi$ is zero, both for linear and logistic models. Namely, even if the IV is invalid $\alpha^* = 0.5$, and there is no causal effect, that is, $\psi = 0$, the G-estimator of $\psi$ is consistent for $\alpha = 0.5$, and its 95%-level CI attains approximately its nominal coverage rate.
3. In the linear causal model, the G-estimators are quite stable with narrow CIs. As a consequence, the coverage probability of the CI depends heavily on the assumed value of $\alpha$. In the examined scenarios, a deviation of 0.1 from the true value of $\alpha^* \in \{0, 0.5\}$ resulted in empirical coverage rates of less than 70% for the nominal 95%-level CIs. On the other hand, for correctly specified $\alpha$, the G-estimator is stable with reliable CIs that achieve the nominal coverage rate.

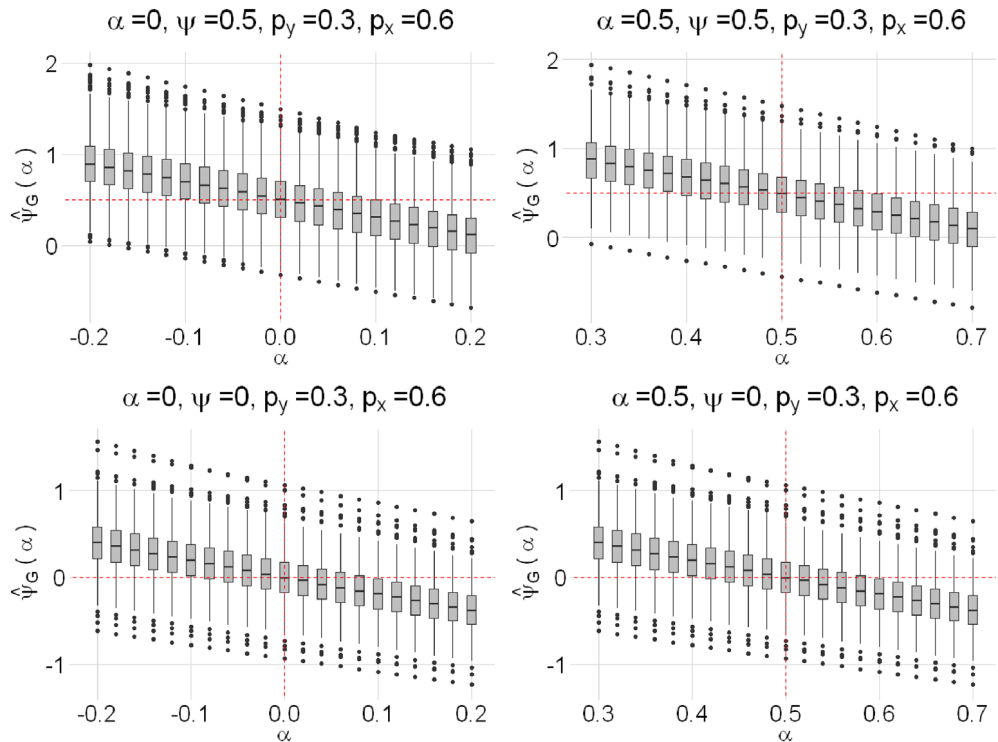

**FIGURE 7** Boxplots of empirical distribution of the G-estimators $\hat{\psi}_G(\alpha)$ of the true causal parameter $\psi \in \{0, 0.5\}$ as a function of the sensitivity parameter $\alpha$, for $\alpha^* \in \{0, 0.5\}$, for sample size $n = 1000$, and $m = 1000$ repetitions. The logistic outcome model $E[Y|X, Z; \beta_Y]$ is as given in (22), $P(Y = 1) = p_y = 0.3$, and $P(X = 1) = p_x = 0.6$. The dashed horizontal red line denotes the true value of $\psi$. The dashed vertical red line denotes the true value of $\alpha$.

4. In the logistic causal model with logistic outcome model, the variance of the G-estimator depends on the dimension and the stability of $\beta_Y$ estimators. Therefore, compared to the linear causal models, the CIs are significantly wider.

5. In the logistic causal model, the simulation is sensitive to the sample size and the specified values of the marginal probability of the outcome $Y$. For example, values of $p_y$ in the vicinity of 0 might require a very large sample size to obtain a solution for the estimating equations and the covariance matrix. A singular covariance matrix might result from relatively low empirical frequencies of every possible combination of $Z, X$, and $Y$, which are crucial for covariance matrices computations. Linear causal models, due to the relative simplicity of their covariance matrices, do not share this property. Therefore, in linear causal models, G-estimators and their corresponding CIs are less sensitive to the sample size and extreme values.

## 4 | REAL WORLD DATA EXAMPLE

In order to illustrate our novel method of sensitivity analysis in a real-world scenario, we use the vitamin D data that are available in the ivtools R-package. These publicly available data are modified version of the original data from a cohort[33] study on vitamin D status causal effect on mortality rates that were previously used by Sjölander & Martinussen.[22] Vitamin D deficiency has been linked with several lethal conditions such as diabetes, cancer, and cardiovascular diseases. However, vitamin D status is also associated with several behavioral and environmental factors, such as season and smoking habits, that may result in biased estimators when using standard statistical analyses to estimate causal effects.

Mendelian randomization[34,35] is a method whose principles were introduced originally by Katan[36] in a strictly medical context. Subsequently, Youngman et al[37] introduced this method in the context of epidemiological studies and also coined the aforementioned term. Mendelian randomization is a method that uses genotypes as IVs to estimate the causal effect of phenotype on disease-related outcomes. The population distribution of genetic variants is assumed to be independent of behavioral and environmental factors that usually confound the effect of exposure on the outcome. The process governing the distribution of genetic variants in the population resembles the randomization mechanism in RCTs. Namely, assuming that this process satisfies the assumptions presented in the Introduction section, the genotype distribution constitutes a form of natural experiment that allows to identify the causal effect of exposure on the outcome. However, these assumptions can be violated, for example, by possible developmental changes that compensate for genetic variations and linkage disequilibrium. Linkage disequilibrium is a term that describes the departure of genetic variants distribution from the independence of confounding factors. Such departure corresponds to a violation of the exogeneity assumption. Developmental changes (also known as Canalization) refer to a situation where the effects of the genotype on the outcome bypass the exposure (phenotype), thus violating the exclusion assumption that the IV affects the outcome only through the exposure. For a detailed discussion of additional possible violations of the IV assumptions in the context of Mendelian randomization, please refer to Lawlor et al.[38] Any combination of these and other possible violations motivates a sensitivity analysis of the causal effect estimator.

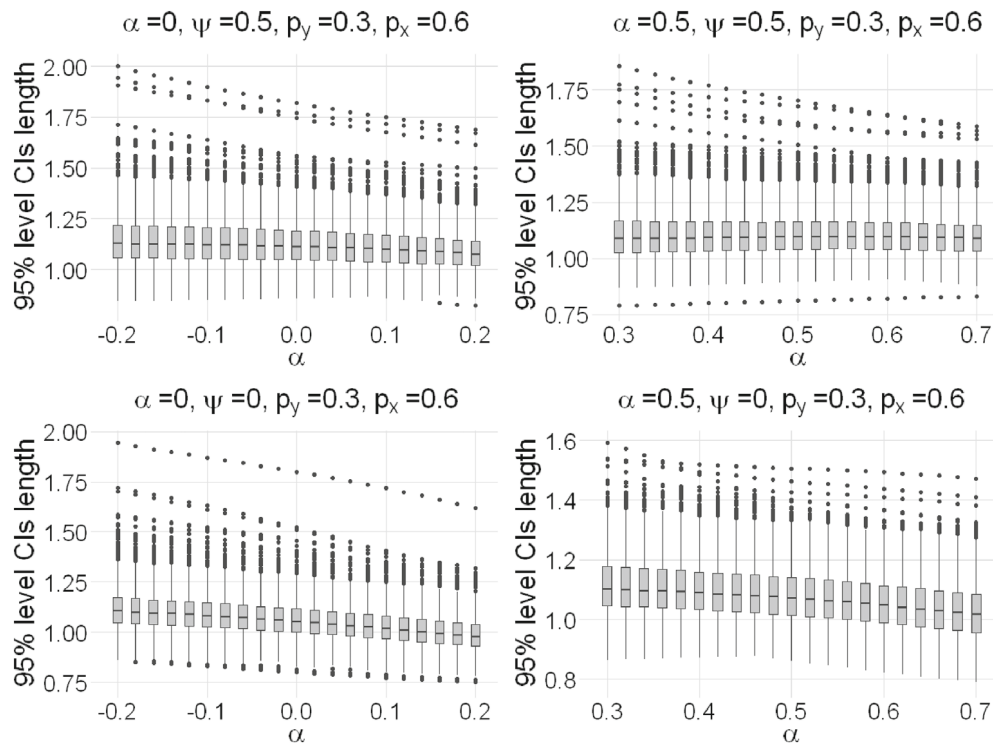

**FIGURE 8** Boxplots of the 95%-level CIs length for $\psi \in \{0, 0.5\}$ in the logistic causal model as a function of the sensitivity parameter $\alpha$, for $\alpha^* \in \{0, 0.5\}$, for sample size $n = 1000$, and $m = 1000$ repetitions. The logistic outcome model for $E[Y|X, Z; \beta_Y]$ is as given in (22), $P(Y = 1) = p_y = 0.3$, and $P(X = 1) = p_x = 0.6$.

In our example, the phenotype is vitamin D status, the diseases related outcome is death during follow-up, and the genotype is mutations in the filaggrin gene. These mutations have been shown to be associated with a higher serum vitamin D concentration. The prevalence of this mutation is estimated to be 8%–10% in the northern European population. Possible developmental changes that compensate for genetic variations, linkage disequilibrium between filaggrin genotype, and possible epigenetic effects[35,39] may violate the IV assumptions of the filaggrin genotype. We used the modified version of data obtained in the Monica10 population-based study. This is a 10-years follow-up study started in 1982–1984 that initially included examination of 3,785 individuals of Danish origin. The participants were recruited from the Danish Central Personal Register as a random sample of the population. In the follow-up study of 1993–1994, the participation rate was about 70% and resulted in a data that contained a total of 2,656 participants, where 2,571 were also available in the modified data after the removal of cases that had missing information on filaggrin and or on vitamin D status.[29,33] This data consisted of 5 variables: age (at baseline), filaggrin (a binary indicator of whether filaggrin mutations are present), vitd (vitamin D level at baseline as was assessed by serum 25-hydroxyvitamin D 25-OH-D(nmol/L) concentration on serum lipids), time (follow-up time), and death (an indicator of whether the subject died during follow-up).

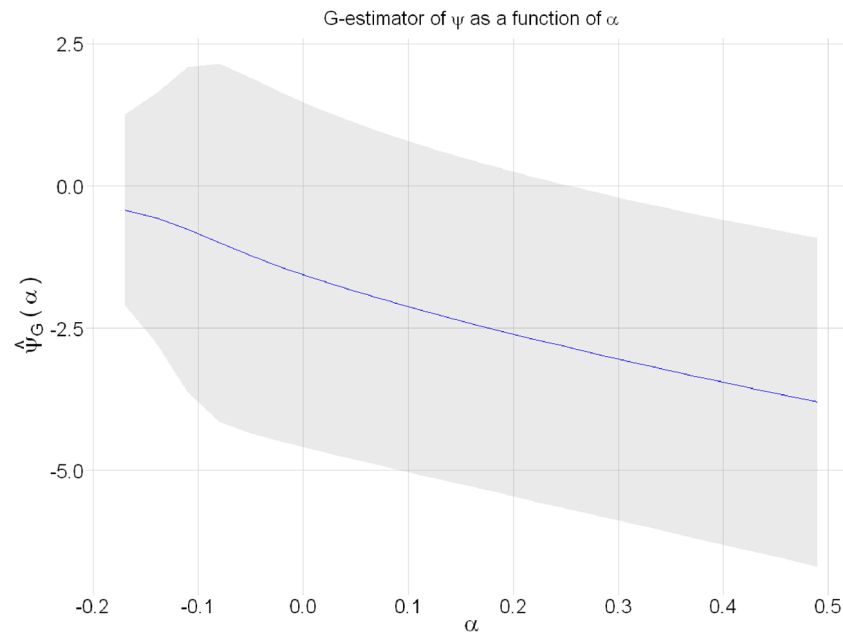

**FIGURE 9**    G-estimator $\hat{\psi}_G(\alpha)$ of the vitamin D baseline status causal effect $\psi$ on death rate during follow-up as a function of the sensitivity parameter $\alpha$. The IV is taken to be the binary indicator of the presence of filaggrin gene mutations. The solid blue line is the point G-estimator. The light gray ribbon is the 95%-level confidence interval based on the G-estimator.

This analysis is mainly for illustration purposes; therefore, we consider only the necessary variables for the required models. We use the binary indicator of death during follow-up as the point outcome $Y$, the binary indicator of the presence of the filaggrin gene mutation as the IV $Z$, and following Martinussen et al[29] the scaled version of the vitamin D status at baseline (vitd) as a continuous exposure variable $X$. The structural equations are as given in Equation (22), and the structural mean model is as given in Equation (17) where the target parameter $\psi$ is the natural logarithm of the causal OR. The marginal probability $p_z$ is estimated with an intercept-only logistic regression model. The relevance assumption, that is, whether the instrument is associated with the exposure, was assessed with a linear model that yielded an F-statistic of 7.349 with 1 and 2569 degrees of freedom. The point estimator of the regression coefficient of filaggrin gene mutation was 0.273 with a corresponding 95%-level CI of [0.076, 0.471]. This result strengthened the legitimacy of the instrument.

To the best of the authors' knowledge, the main concern in our setting is the violation of the exogeneity assumption.[40,41] Namely, possibly unmeasured confounders associated with the IV filaggrin mutation and death during follow-up. In such a case, the sensitivity parameter $\alpha$ represents a possible bias of the estimated causal effect (w.r.t. the true causal effect) induced by the unmeasured confounders of the outcome and the IV. According to the vast clinical literature on the association between vitamin D deficiency and mortality rates, the true causal effect is expected to be negative.[33,42] The exact mechanism by which the vitamin D affects the mortality rates is yet unknown;[41] thus, we cannot rule out the possibility of no causal effect. However, since there is no known clinical evidence of the positive effects of vitamin D deficiency on mortality rates, in the sensitivity analysis, we can discard values of $\alpha$ that correspond to positively valued estimators of $\psi$. We can also discard values of $\alpha$ that correspond to unprecedentedly high causal effects with no previous records or plausible clinical explanation.[43] Therefore, we empirically derive the set of plausible values of $\alpha$ to be considered in the sensitivity analysis. The G-estimator $\hat{\psi}_G(\alpha)$ of the vitamin D causal effect $\psi$ on death during follow-up as a function of the sensitivity parameter $\alpha$ is illustrated in Figure 9, and in a corresponding Table 3.

If the filaggrin gene mutation is a valid IV, that is, assuming $\alpha^* = 0$, the G-estimator of $\psi$ equals 1.558 with a corresponding 95%-level CI spanning the 0 value. Additionally, there is no solution to the estimating equations for $\alpha < -0.17$. Such results support the possibility of a bias; an underestimation of the magnitude of the true causal effect of vitamin D deficiency on mortality rate during follow-up. For all considered values of $\alpha$, the sign of the point G-estimator does not change (see Table 3 for further details). Therefore, we may conclude that if the structural mean models and the violation structure are correctly specified (given that the data constitutes a representative sample of the target population and the sample variability can be neglected), there is no evidence of a positive causal effect of vitamin D deficiency on the

**TABLE 3** G-estimator of $\psi$ and OR, with the corresponding 95%-level CI limits, of the vitamin D baseline status effect on death rate during follow-up as a function of $\alpha$.

| $\alpha$ | $\hat{\psi}(\alpha)$ | 95% level CI | $\hat{OR} = e^{\hat{\psi}(\alpha)}$ | 95% level CI |
|---|---|---|---|---|
| −0.15 | −0.512 | [−2.502, 1.478] | 0.599 | [0.082, 4.382] |
| −0.10 | −0.842 | [−3.852, 2.168] | 0.431 | [0.021, 8.742] |
| −0.05 | −1.224 | [−4.349, 1.902] | 0.294 | [0.013, 6.698] |
| **0.00** | **−1.558** | **[−4.588,1.472]** | **0.211** | **[0.010, 4.358]** |
| 0.05 | −1.853 | [−4.811, 1.105] | 0.157 | [0.008, 3.020] |
| 0.10 | −2.121 | [−5.028, 0.786] | 0.120 | [0.007, 2.195] |
| 0.15 | −2.369 | [−5.241, 0.503] | 0.094 | [0.005, 1.654] |
| 0.20 | −2.603 | [−5.453, 0.248] | 0.074 | [0.004, 1.282] |
| 0.25 | −2.825 | [−5.666, 0.015] | 0.059 | [0.003, 1.015] |
| 0.30 | −3.040 | [−5.878, −0.201] | 0.048 | [0.003, 0.818] |
| 0.35 | −3.247 | [−6.091, −0.403] | 0.039 | [0.002, 0.668] |
| 0.40 | −3.449 | [−6.304, −0.594] | 0.032 | [0.002, 0.552] |
| 0.45 | −3.646 | [−6.517, −0.775] | 0.026 | [0.001, 0.461] |
| 0.50 | −3.840 | [−6.731, −0.948] | 0.022 | [0.001, 0.387] |

*Note*: The row in bold denotes the estimators for $\alpha = 0$, that is, when no violation is assumed.

mortality rate. Particularly, for $\alpha = -0.17$, the G-estimator $\hat{\psi}(\alpha) = -0.42$ with 95%-level CI [−0.42, 2.10]. In other words, the absence of a solution implies that values of $\alpha$ below −0.17 are logically impossible (ie, incompatible with) for these particular data. Namely, under the assumptions above, the nonexistence of solution for $\alpha < -0.17$ provides an upper bound of the vitamin D deficiency causal effect on mortality rates during follow-up.[44] If the IV is invalid, the true effect of vitamin D deficiency on the mortality rate is likely to be of a larger magnitude than the estimated value for $\alpha = 0$. Such findings are consistent with the clinical literature on the negative effects of vitamin D deficiency.

## 5 | SUMMARY & CONCLUSIONS

In this study, we propose and demonstrate a new sensitivity analysis method for G-estimators in structural mean models. This study introduces two novel aspects of sensitivity analysis. The first is a single sensitivity parameter that captures violations of the exclusion and exogeneity assumptions. Using a single parameter is a valuable advantage of the new method as fewer parameters are needed to be specified. This feature may increase model stability and decrease the risk of misspecification. The second is the application of the method to non-linear models. The proposed method is theoretically justified and is illustrated via a simulation study and a real-world example. This study highlights the importance of sensitivity analysis for G-estimators of causal parameters and provides general guidelines for conducting such analysis. However, this study is not without limitations. The usage of one parameter for two distinct violations complicates its interpretation. Therefore, specifying a plausible interval for the sensitivity parameter may be challenging since it relies heavily on subject-matter knowledge and a solid understanding of causal inference. Another limitation is of a computational nature. This study examines quite limited scenarios in the simulation study. For example, we considered only binary exposure with a binary instrument in the logistic causal model. Although considering continuous instruments in the logistic causal model introduces computational complications, it should not induce any qualitative difference to the proposed sensitivity analysis. An additional computational limitation is that the estimating equations were solved only for a single instrumental variable. This study's main contribution is providing a theoretical framework for conducting sensitivity analysis for G-estimators. The presented framework can be readily extended to apply to various structural mean models and computationally involved scenarios, which may serve as directions for future research.

## DATA AVAILABILITY STATEMENT

The source code of the simulations that support the findings of this study (Section 3) is openly available in GitHub repository https://github.com/vancak/G_sensitivity. The data that support the findings of Section 4 are openly available as part of the ivtools R-package in CRAN repository at https://CRAN.R-project.org/package=ivtools.

## ENDNOTES

*For non-linear target parameters, for example, OR and hazard ratio (HR), more careful analysis is required even in RCTs. This is due to the non-collapsibility of these measures. For example, in Cox models, the implicit conditioning of the risks set on their survival up to a certain time point challenges the causal interpretation of the HR estimators. However, since in this study, we are not targeting marginal OR or HR in survival models, these issues are outside the scope of this paper. For further details, we refer the reader to Aalen et al. (2015).[1]

†The simulations, figures and Vitamin D data analysis source code: https://github.com/vancak/G_sensitivity

## ORCID

*Valentin Vancak*  https://orcid.org/0000-0001-8732-7353

*Arvid Sjölander*  https://orcid.org/0000-0001-5226-6685

---

## APPENDIX

### Violations structure equivalence

Following Wang et al[13] notations, let $Y^{(x,z)}$ be the potential outcome when the exposure $X$ is set to $x$ and the IV $Z$ to $z$. According to Wang et al,[13] the direct causal effect of the IV $Z$ on the outcome $Y$, $Z \to Y$, is assumed to be homogeneous across individuals and is defined as

$$Y^{(0,z)} - Y^{(0,0)} = \delta_1 z. \tag{A1}$$

Notably, definition (A1) implies the following form (A1ba) of the exclusion assumption violation

$$E[Y^{(0,Z)} - Y^{(0,0)}|Z] = \delta_1 Z. \tag{A1ba}$$

In addition, Wang et al defined the induced association between the IV $Z$ and the outcome $Y$, $Z \longleftarrow U \longrightarrow Y$, by unmeasured confounders, as

$$E[Y^{(0,0)}|Z = z] - E[Y^{(0,0)}|Z = 0] = \delta_2 z.$$

In this article, for a link function $\xi$ of a generalized linear model, we define the assumptions' violation as $\xi(E[Y_0|Z]) - \xi(E[Y_0|Z = 0])$. The following derivation shows the equivalence between the two definitions for a link function $\xi$

$$\xi(E(Y_0|Z)) - \xi(E(Y_0|Z=0))$$
$$= \xi(E(Y^{(0,Z)}|Z)) - \xi(E(Y^{(0,0)}|Z=0))$$
$$= \xi(E(Y^{(0,Z)}|Z)) - \xi(E(Y^{(0,0)}|Z)) + \xi(E(Y^{(0,0)}|Z)) - \xi(E(Y^{(0,0)}|Z=0))$$
$$= \xi(E(Y^{(0,Z)}|Z)) - \xi(E(Y^{(0,0)}|Z)) + \left(\xi(E(Y^{(0,0)}|Z) - \xi(E(Y^{(0,0)}|Z=0))\right)$$
$$= \delta_1 Z + \delta_2 Z$$
$$= \delta Z. \tag{A2}$$

Using our notations we denote $\delta \equiv \alpha^*$. Wang et al. only considered a model with an identity link function $\xi$. However, a straight-forward generalization of their definitions to the logit link function allows to express both violations on the logit scale. In such case, the direct causal effect of the IV $Z$ on the outcome $Y$ is defined as

$$\text{logit} E(Y^{(0,Z)}|Z) - \text{logit} E(Y^{(0,0)}|Z) = \delta_1 Z,$$

and the induced association between the IV $Z$ and the outcome $Y$ by unmeasured confounders is defined as

$$\text{logit} E(Y^{(0,0)}|Z) - \text{logit} E(Y^{(0,0)}|Z=0) = \delta_2 Z.$$

Using the same derivation as in Equation (A2) for the logit link function and our definition of the assumptions' violation, we can express both violations with a single parameter $\delta$

$$\text{logit} E(Y_0|Z) - \text{logit} E(Y_0|Z=0) = \delta Z,$$

where, in our notations, $\delta \equiv \alpha^*$.

**Proof of Lemma 1**

*Proof.* When $Y$ is a point outcome and $\xi$ is the identity link function, we have that

$$E\{h(\psi; \alpha^*)|L, Z\} = E[E\{h(\psi; \alpha^*)|L, Z, X\}L, Z]$$
$$= E[E\{Y - m^T(L)X\psi - b(L, Z; \alpha^*)|L, Z, X\}|L, Z]$$
$$= E\{E(Y_0|L, Z, X) - b(L, Z; \alpha^*)|L, Z\}$$
$$= E(Y_0|L, Z) - b(L, Z; \alpha^*)$$
$$= a(L),$$

where the third equality follows from Equations (14) and (8).  ∎

**Proof of Lemma 2**

For a linear causal model as in Equation (14) with $L = \emptyset$, and $m(L) = 1$, the unbiased estimating equation for $\psi$ is

$$E[(Z - \mu_Z)(Y - \psi X - \alpha^* Z)] = 0,$$

then

$$0 = E[Y(Z - \mu_Z) - \psi E[X(Z - \beta_Z) - \alpha^* E[Z(Z - \mu_Z)]$$
$$= cov(Y, Z) - \psi cov(X, Z) - \alpha^* var(Z),$$

re-arranging the equation, we get

$$\psi = \frac{cov(Y, Z)}{cov(X, Z)} - \frac{var(Z)}{cov(X, Z)}$$
$$= \frac{\beta_{YZ}}{\beta_{XZ}} - \frac{\alpha^*}{\beta_{XZ}}.$$

**Proof of Lemma 3**

The asymptotic bias of the G-estimator for invalid instrument with $b(L, Z; \alpha^*) = \alpha^* Z$.

*Proof.* Assume a linear structural mean model as described in (14), and that there are no unmeasured confounders, that is, $U = \emptyset$. The G-estimator equals the TSLS estimator.[22] Namely, the G-estimator converges in probability to

$$
\begin{aligned}
\psi_G &= \frac{cov(Y, Z|L)}{cov(X, Z|L)} \\
&= \frac{cov(a(L) + \psi X + \beta_{YZ.L} Z, Z|L)}{cov(X, Z|L)} \\
&= \psi \frac{cov(X, Z|L)}{cov(X, Z|L)} + \beta_{YZ.L} \frac{\sigma^2_{Z.L}}{cov(X, Z|L)} \\
&= \psi + \frac{\beta_{YZ.L}}{\beta_{XZ.L}}.
\end{aligned}
$$

Consequently, the asymptotic bias of the estimator is $\beta_{YZ.L}/\beta_{XZ.L}$. ∎

**Proof of Lemma 4**

In a linear causal model without unmeasured confounders, the corrected G-estimator equals the least squares estimator of $\psi$.

*Proof.* In a linear structural mean model as in (14), the ordinary least squares estimator of $\psi$ converges asymptotically to

$$
\begin{aligned}
\frac{cov(Y, X|L)}{\sigma^2_{X.L}} &= \frac{cov(a(L) + \psi X + \beta_{YZ.L} Z, X|L)}{\sigma^2_{X.L}} \\
&= \psi + \beta_{YZ.L} \frac{cov(Z, X|L)}{\sigma^2_{X.L}} \\
&= \psi + \beta_{YZ.L} \beta_{ZX.L},
\end{aligned}
$$

where for a linear model of $X$ as a function of $Z$ we have that $\beta_{ZX.L} = 1/\beta_{XZ.L}$. Therefore,

$$
\begin{aligned}
\frac{cov(Y, X|L)}{\sigma^2_{X.L}} &= \psi + \frac{\beta_{YZ.L}}{\beta_{XZ.L}} \\
&= \psi_G,
\end{aligned}
$$

such that $\psi_G$ is the asymptotic value of the G-estimator as in Lemma 3. ∎

**Proof of Theorem 1**

*Proof.* In order to estimate $\alpha^*$ and a one-dimensional parameter $\psi$, a system of two estimating equations is required. We construct another function $D_2 = D_2(L, Z)$ that is linearly independent of $D_1(L, Z)$, and satisfies $E[D_2(L, Z)|L] = 0$. Hence, in a linear causal model, we can construct the following sub-system of estimating equations

$$
E \begin{pmatrix} D_1(L, Z) h(\psi; \alpha) \\ D_2(L, Z) h(\psi; \alpha) \end{pmatrix} = E \begin{pmatrix} D_1(L, Z)(Y - X\psi - \alpha Z) \\ D_2(L, Z)(Y - X\psi - \alpha Z) \end{pmatrix} = 0.
$$

Note that $E[D_i Y] = cov(D_i, Y)$ since $E[D_i(Z, L)] = E[E[D_i(Z, L)|L]] = 0$, for $i = 1, 2$. Therefore,

$$
E[D_i(L, Z)(Y - X\psi - \alpha Z)] = cov(D_i, Y) - \psi cov(D_i, X) - \alpha cov(D_i, Z) = 0,
$$

for $i = 1, 2$. Namely, the two bottom rows of the system of estimating equations result in the following sub-system

$$
\begin{pmatrix} cov(D_1, X) & cov(D_1, Z) \\ cov(D_2, X) & cov(D_2, Z) \end{pmatrix} \begin{pmatrix} \psi \\ \alpha \end{pmatrix} = \begin{pmatrix} cov(D_1, Y) \\ cov(D_2, Y) \end{pmatrix},
$$

where for a valid IV $Z$ the coefficients matrix is singular. Therefore, there is a degree of freedom which is the sensitivity parameter $\alpha$. ∎

## Simulation study

### Linear causal model–Simulation setup

The structural mean model is

$$
E[Y_x | X = x, Z] - E[Y_0 | X = x, Z] = \psi x,
$$

and the violation is defined as

$$
E[Y_0 | Z] - E[Y_0 | Z = 0] = \alpha^* Z.
$$

Assume that both $X$ and $Z$ are binary. We use the DGP presented in (20) to express $E[Y_0 | Z]$ as a function of the observed data and the causal parameter $\psi$. Then we use the DGP presented in (20) and the violation structure to obtain the functional relationship between the sensitivity parameter $\alpha^*$ and the DGP parameters of $Y$

$$
\begin{aligned}
E[Y_0 | Z] &= E_X[E[Y_0 | Z, X]] \\
&= E[Y_0 | X = 0, Z] P(X = 0 | Z) + E[Y_0 | X = 1, Z] P(X = 1 | Z) \\
&= (E[Y | X = 0, Z]) P(X = 0 | Z) + (E[Y | X = 1, Z] - \psi) P(X = 1 | Z) \\
&= (\beta_0 + \beta_z Z)(1 - P(X = 1 | Z)) + (\beta_0 + \beta_x + \beta_z Z + \beta_{xz} Z - \psi) P(X = 1 | Z) \\
&= (\beta_0 + \beta_z Z)(1 - \text{expit}(\gamma_0 + \gamma_z Z)) + (\beta_0 + \beta_x + \beta_z Z + \beta_{xz} Z - \psi) \text{expit}(\gamma_0 + \gamma_z Z).
\end{aligned}
$$

Therefore,

$$
\begin{aligned}
E[Y_0 | Z = 1] &- E[Y_0 | Z = 0] \\
&= (\beta_0 + \beta_z)(1 - \text{expit}(\gamma_0 + \gamma_z)) + (\beta_0 + \beta_x + \beta_z + \beta_{xz} - \psi) \text{expit}(\gamma_0 + \gamma_z) \\
&- \beta_0 (1 - \text{expit}(\gamma_0)) - (\beta_0 + \beta_x - \psi) \text{expit}(\gamma_0) \\
&= \alpha^*.
\end{aligned}
$$

### Logistic causal model–Simulation setup

The structural mean model is defined on the logit scale

$$
\text{logit} P(Y_x = 1 | X = x, Z) - \text{logit} P(Y_0 = 1 | X = x, Z) = \psi x.
$$

The violation is defined on the logit scale as well

$$
\text{logit} P(Y_0 = 1 | Z) - \text{logit} P(Y_0 = 1 | Z = 0) = \alpha^* Z.
$$

Assume that both $X$ and $Z$ are binary. We use the DGP presented in (22) to express $P(Y_0 = 1 | Z)$ as a function of the observed data and the causal parameter $\psi$. Then we use the DGP presented in (22) and the violation structure to obtain the functional relationship between the sensitivity parameter $\alpha^*$ and the DGP parameters of $Y$

$$
\begin{aligned}
&P(Y_0 = 1|Z) \\
&= P(Y_0 = 1|X = 0, Z)P(X = 0|Z) + P(Y_0 = 1|X = 1, Z)P(X = 1|Z) \\
&= P(Y = 1|X = 0, Z)P(X = 0|Z) + P(Y_0 = 1|X = 1, Z)P(X = 1|Z) \\
&= P(Y = 1|X = 0, Z)P(X = 0|Z) + \text{expit}(\text{logit}P(Y_0 = 1|X = 1, Z))P(X = 1|Z) \\
&= P(Y = 1|X = 0, Z)P(X = 0|Z) + \text{expit}(\text{logit}P(Y_1 = 1|X = 1, Z) - \psi)P(X = 1|Z) \\
&= P(Y = 1|X = 0, Z)P(X = 0|Z) + \text{expit}(\text{logit}P(Y = 1|X = 1, Z) - \psi)P(X = 1|Z) \\
&= \text{expit}(\beta_0 + \beta_z Z)(1 - \text{expit}(\gamma_0 + \gamma_z Z)) + \text{expit}(\beta_0 + \beta_x + \beta_z Z + \beta_{xz} Z - \psi)\text{expit}(\gamma_0 + \gamma_z Z).
\end{aligned}
$$

Therefore,

$$
\begin{aligned}
&\text{logit}P(Y_0 = 1|Z = 1) - \text{logit}P(Y_0 = 1|Z = 0) \\
&= \text{logit}(\text{expit}(\beta_0 + \beta_z)(1 - \text{expit}(\gamma_0 + \gamma_z)) + \text{expit}(\beta_0 + \beta_x + \beta_z + \beta_{xz} - \psi)\text{expit}(\gamma_0 + \gamma_z)) \\
&\quad - \text{logit}(\text{expit}(\beta_0)(1 - \text{expit}(\gamma_0)) + \text{expit}(\beta_0 + \beta_x - \psi)\text{expit}(\gamma_0)) \\
&= \alpha^*.
\end{aligned}
$$

We have nine unknown parameters $\theta_0^T = (\beta_0, \beta_x, \beta_z, \beta_{xz}, \gamma_0, \gamma_z, \psi, \alpha, p_z)$ and the sensitivity parameter $\alpha$. If we specify the causal effect $\psi$, and the marginal probabilities of $Z = 1$, $X = 1$, and $Y = 1$, we left with four degrees of freedom.

**Logistic causal model–Simulation results**

Table A1 and Figures A1, A2, A3 here.

**TABLE A1** Logistic causal model: Mean values of the 95%-level CIs of $\psi = 0, 0.5$ as a function of $\alpha$.

| $\alpha$ | $\psi = 0$ | $\psi = 0.5$ | $\alpha$ | $\psi = 0$ | $\psi = 0.5$ |
| --- | --- | --- | --- | --- | --- |
| −0.2 | 0.86 | 0.972 | 0.3 | 0.912 | 1.043 |
| −0.18 | 0.857 | 0.967 | 0.32 | 0.908 | 1.039 |
| −0.16 | 0.853 | 0.963 | 0.34 | 0.905 | 1.035 |
| −0.14 | 0.849 | 0.958 | 0.36 | 0.901 | 1.031 |
| −0.12 | 0.846 | 0.953 | 0.38 | 0.898 | 1.027 |
| −0.1 | 0.843 | 0.949 | 0.4 | 0.895 | 1.023 |
| −0.08 | 0.839 | 0.944 | 0.42 | 0.891 | 1.019 |
| −0.06 | 0.836 | 0.94 | 0.44 | 0.888 | 1.015 |
| −0.04 | 0.833 | 0.936 | 0.46 | 0.885 | 1.01 |
| −0.02 | 0.83 | 0.931 | 0.48 | 0.882 | 1.006 |
| 0 | 0.828 | 0.927 | 0.5 | 0.88 | 1.002 |
| 0.02 | 0.825 | 0.923 | 0.52 | 0.877 | 0.998 |
| 0.04 | 0.822 | 0.918 | 0.54 | 0.874 | 0.994 |
| 0.06 | 0.82 | 0.914 | 0.56 | 0.872 | 0.99 |
| 0.08 | 0.818 | 0.91 | 0.58 | 0.869 | 0.986 |
| 0.1 | 0.815 | 0.906 | 0.6 | 0.867 | 0.982 |
| 0.12 | 0.813 | 0.902 | 0.62 | 0.865 | 0.978 |
| 0.14 | 0.811 | 0.898 | 0.64 | 0.863 | 0.974 |
| 0.16 | 0.809 | 0.895 | 0.66 | 0.861 | 0.97 |
| 0.18 | 0.807 | 0.891 | 0.68 | 0.859 | 0.966 |
| 0.2 | 0.805 | 0.887 | 0.7 | 0.857 | 0.963 |

*Note*: The true $\alpha^* = 0, 0.5$, respectively. The marginal distribution f the outcome is $P(Y = 1) = p_y = 0.8$, and $P(X = 1) = p_x = 0.6$.

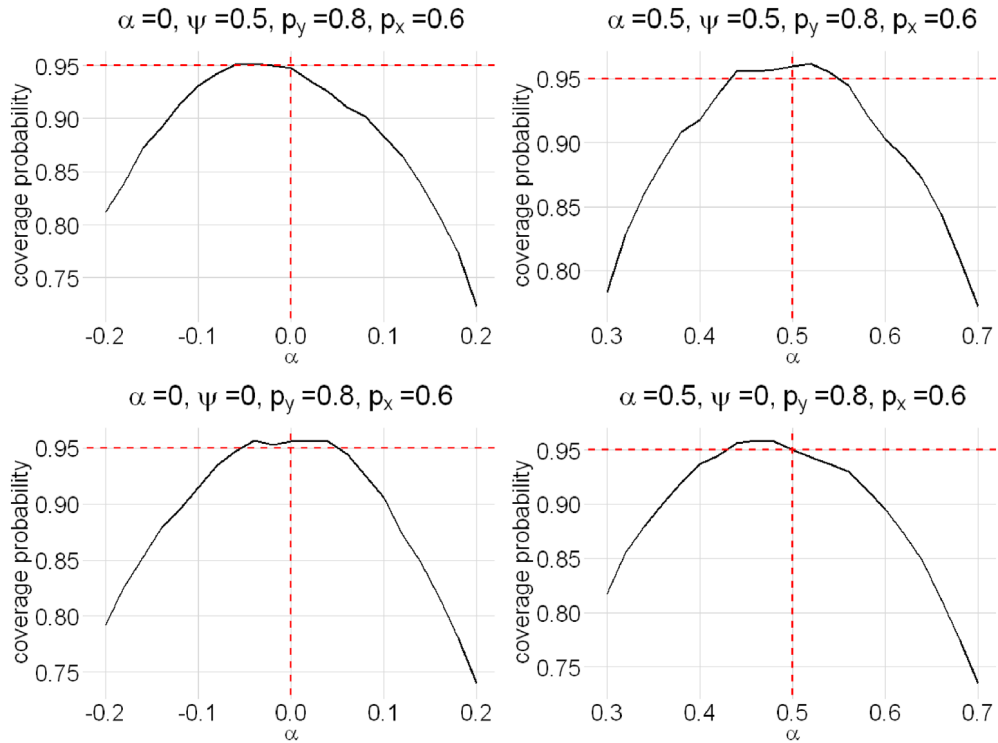

**FIGURE A1**  Coverage rates of the 95%-level CIs of the true causal parameter $\psi \in \{0, 0.5\}$ in the logistic causal model as a function of the sensitivity parameter $\alpha$, for $\alpha^* \in \{0, 0.5\}$, for sample size $n = 1000$, and $m = 1000$ repetitions. The logistic outcome model $E[Y|X, Z; \beta_Y]$ is as given in (22), and $P(Y = 1) = p_y = 0.8$. The dashed red horizontal line denotes the 95% coverage rate. The solid black line denotes the computed coverage probabilities.

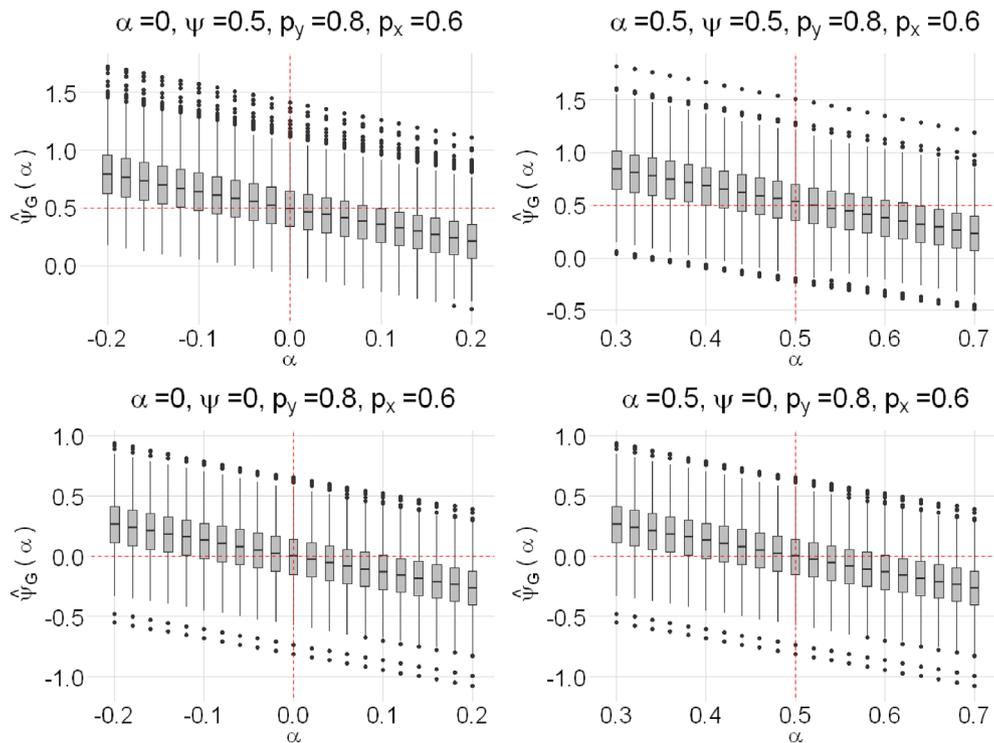

**FIGURE A2**  Boxplots of the empirical distribution of G-estimators $\hat{\psi}_G(\alpha)$ of the true causal parameter $\psi \in \{0, 0.5\}$ in the logistic causal model as a function of the sensitivity parameter $\alpha$, for $\alpha^* \in \{0, 0.5\}$, for sample size $n = 1000$, and $m = 1000$ repetitions. The logistic outcome model $E[Y|X, Z; \beta_Y]$ is as given in (22), and $P(Y = 1) = p_y = 0.8$. The dashed horizontal red line denotes the true value of $\psi$. The dashed vertical red line denotes the true value of $\alpha$.

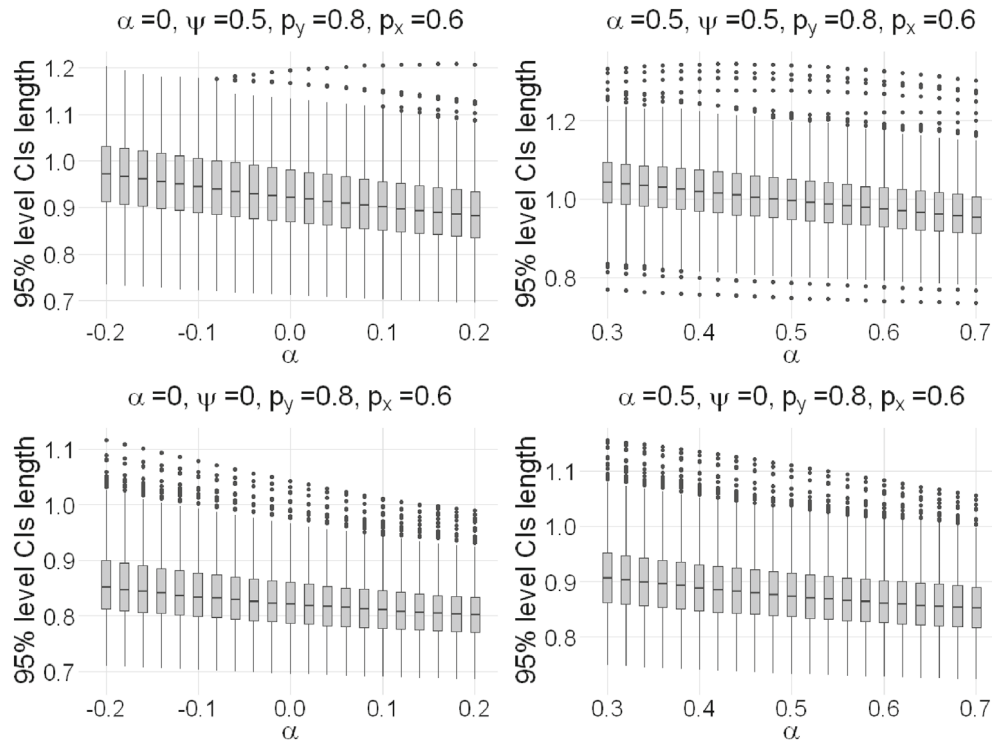

**FIGURE A3**  Boxplots of the 95%-level CIs length for $\psi \in \{0, 0.5\}$ in the logistic causal model as a function of the sensitivity parameter $\alpha$, for $\alpha^* \in \{0, 0.5\}$, for sample size $n = 1000$, and $m = 1000$ repetitions. The logistic outcome model for $E[Y|X, Z; \beta_Y]$ is as given in (22), $P(Y = 1) = p_y = 0.8$, and $P(X = 1) = p_x = 0.6$.